\documentclass[
,preprint
,secnumarabic
,amssymb, amsmath, nobibnotes, aps, prl, 
]{revtex4}

\usepackage{graphicx}
\usepackage[usenames]{color}

\begin{document}

\title{Halving the Casimir force with conductive oxides: experimental details}

\author{S.~de~Man}
\author{K.~Heeck}
\author{D.~Iannuzzi}
\email{iannuzzi@few.vu.nl}
\affiliation{Department of Physics and Astronomy, VU University
Amsterdam, De Boelelaan 1081, 1081 HV Amsterdam, The Netherlands}

\date{\today}

\begin{abstract}
This work is an extended version of a paper published last year in Physical Review Letters [S.~de~Man~\emph{et al.}, Phys.~Rev.~Lett.~\textbf{103}, 040402 (2009)], where we presented measurements of the Casimir force between a gold coated sphere and a plate coated with either gold or an indium-tin-oxide (ITO) layer. The experiment, which was performed in air, showed that ITO is sufficiently conducting to prevent charge accumulation, but still transparent enough to halve the Casimir attraction when compared to gold. Here, we report all the experimental details that, due to the limited space available, were omitted in the previous article. We discuss the performance of our setup in terms of stability of the calibration procedure and reproducibility of the Casimir force measurement. We also introduce and demonstrate a new technique to obtain the spring constant of our force sensor. Furthermore, we present a thorough description of the experimental method, a comprehensive explanation of data elaboration and error analysis, and a complete characterization of the dielectric function and of the surface roughness of the samples used in the actual experiment.
\end{abstract}

\maketitle

\section{Introduction}
It is well known that the Casimir effect~\cite{Casimir:1948p440} strongly depends on the dielectric function of the interacting surfaces~\cite{lifshitz,parsegian}. Transparent dielectrics, for example, attract less than highly reflective metals. Dielectric materials, however, tend to accumulate isolated charges. Those charges give rise to an electrostatic force that easily overcomes the Casimir interaction.

In a recent paper~\cite{deMan:2009p99}, we have presented measurements of the Casimir force between a gold coated sphere and a plate coated with either gold or an Indium-Tin-Oxide (ITO, In$_2$O$_3$:Sn) layer. The experiment, which was performed in air, showed that ITO is sufficiently conducting to prevent charge accumulation, but still transparent enough to halve the Casimir attraction when compared to gold.

The experiment was carried out by means of a quite complicated novel technique that, due to the limited space available, was not thoroughly explained in our previous work. We believe it is important to extend that work and provide the community with all the details of the experimental technique and data analysis, which is the purpose of this paper.

This paper is organized as follows. First, we describe the experimental setup and discuss general issues one has to tackle to perform Casimir force measurements. Then we discuss the experimental technique we developed to simultaneously calibrate the setup and measure the Casimir force gradient, and derive in detail the specific forms of all our calibration and measurement signals. Second, we illustrate a new method to determine the spring constant of our force sensor. Third, we present experimental results on the general performance of our setup, namely the stability of the calibration procedure, the reproducibility of the force gradient measurements, and the spring constant determination. Fourth, we present the Casimir force measurements for the gold-gold and gold-ITO interactions, and show measurements of the dielectric functions and surface topographies of our surfaces. Finally we compare the hydrodynamic forces for the two sets of experiments.

\section{Experimental setup}

\subsection{Description}
Our experimental setup is designed to precisely measure surface forces between a 100~$\mu$m radius sphere and planar samples at ambient pressure. The sphere is attached to a micromachined cantilever (spring constant roughly 1~N/m) whose deflection in response to external forces can be measured with pm sensitivity by a commercial Atomic Force Microscope (AFM) detection head (Veeco Multimode); the detection system is formed by a laser beam that reflects from the free end of the cantilever and hits a position sensitive photodetector (see Fig.~\ref{fig:schematic}). The sphere-cantilever assembly is coated with a Ti adhesion layer and a 100~nm Au film. The planar sample is mounted on a two-stage mechanical translator formed by a stick-slip piezoelectric motor (Attocube) and a feedback controlled piezoelectric transducer (Physik Instrumente) to vary the separation between the sphere and plate surfaces. The stick-slip motor is used for coarse approach (travel range 6mm), while the feedback controlled transducer executes the fine distance scanning (range 12~$\mu$m, closed loop resolution 50~pm). Both the detection head and the two-stage mechanical translator are anchored to a 10~cm$^{3}$ Al block that is actively temperature stabilized at 300~K to reduce mechanical drift from differential thermal expansion of the components. The Al block is screwed onto an active anti-vibration table (Halcyonics), which is placed inside an anechoic chamber. This chamber lies onto a heavy marble optical table that is located in a temperature controlled laboratory.

\subsection{Three crucial issues}

In a Casimir force measurement, there are three crucial issues that have to be dealt with.

First, even if one would electrically connect both interacting surfaces, there exists an electrostatic potential difference $V_{0}$ due to the different work functions of the surfaces. Since work functions of surfaces depend on quite a number of parameters, like crystal growth orientation and adsorbates, typically there even exists a potential difference between surfaces made out of the same material. This electrostatic potential difference gives rise to a force that is generally stronger than the Casimir force. To avoid this problem, most Casimir force setups rely on a counterbias circuit that is used to apply $-V_{0}$ to the surfaces in order to have no residual electrostatic force.

Second, even in setups where the distance between the sphere and the plate is varied with a feedback controlled piezoelectric transducer, one has only knowledge of the relative position changes and not of the absolute separation between the surfaces. It is thus mandatory to find the initial separation $d_{0}$ with a calibration procedure. Because the distance dependence of the electrostatic force between a sphere and a plate is known exactly, most modern setups use this force to extract $d_{0}$.

Third, the instrument has to be calibrated with a known force. Again, one can use the electrostatic interaction to calibrate photodetector voltage versus force. We have developed a measurement scheme that solves all three issues at the same time.

\subsection{Force modulation measurements}
We present a measurement technique that makes use of simultaneous detection of both calibration signals (based on the electrostatic force) and the Casimir force. The motivation for this approach is the benefit of absolute certainty that the calibration parameters always correspond to the measured forces because they are acquired simultaneously; it is thus impossible to have inconsistent calibration and force data due to time-related drifts or other events. In order to achieve this goal, we have separated the calibration and Casimir signals in frequency space: the signals are modulated at distinct frequencies that can be de-modulated individually with lock-in amplifiers.

\begin{figure}[h!]
\includegraphics[width=13cm]{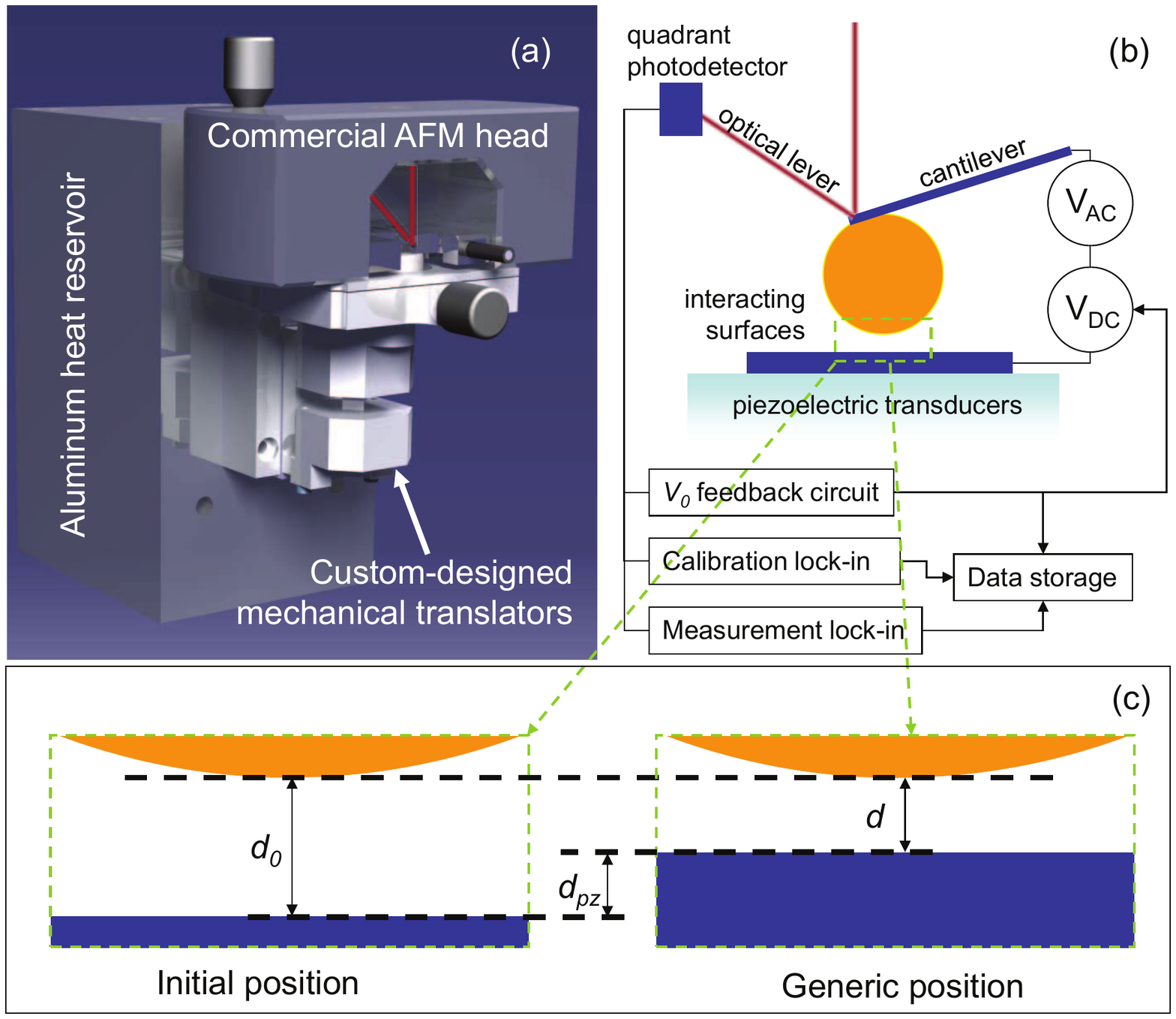}
\caption{(Color online) (a) Drawing of the experimental setup used
to perform precise measurements of the Casimir force between a 100 $\mu$m radius sphere and a
plate. The aluminum block acts as a heat reservoir to keep the temperature of the setup constant.
The instrument is based on a commercial AFM head that is, together with a custom-designed
mechanical translator, mounted on the aluminum block. (b) Schematic representation of the working principle of the
experimental technique. The $V_0$ feedback circuit allows one to measure and compensate the
residual voltage present between the sphere and the plate. The calibration lock-in amplifier is
used to calibrate the instrument and to find the initial separation between the two surfaces $d_0$.
The measurement lock-in amplifier performs the measurements of the Casimir force gradient and the hydrodynamic force. (c) Definition of the initial separation $d_0$, the movement of the feedback controlled piezoelectric stage $d_{pz}$, and the non-modulated separation between the surfaces $d=d_{0}-d_{pz}$ as used in the Taylor expansion that leads to Eqs.~\ref{Stimeseries} to~\ref{Somega2Q}.} \label{fig:schematic}
\end{figure}

Modulating an electrostatic interaction is extremely easy: one just has to apply a time-dependent potential difference to the sphere and the plate ($V_{AC}$ in Fig.~\ref{fig:schematic}~b). We thus apply an oscillating voltage $V_{DC} + V_{AC} \cos \left( \omega_1 t \right)$ between the
sphere and the plate, where $V_{DC}$ is used to compensate for the contact potential difference $V_{0}$~\cite{deMan:2009p104}. Unfortunately, modulating the Casimir force is a lot more challenging as its strength depends only on geometry and dielectric properties of the surfaces. On the other hand, of course, the strong distance dependence of the Casimir force can be used to modulate its strength considerably. Therefore, we chose to add a small modulation of the form $\Delta d \cos\left(\omega_2 t\right)$ to the piezoelectric transducer displacement $d_{pz}$, as previously introduced in \cite{Bressi:2001p110}. When the sphere and plate surfaces are separated by a distance $d$, we have the following three forces acting on the sphere:
$F(V,d,\omega_{2},\Delta d)=F_E(V,d)+F_C(d)+F_H(d,\omega_{2},\Delta d)$ where $F_E(V,d)$ is the electrostatic force for externally applied potential difference $V$, $F_C(d)$ is the
Casimir force, and $F_H(d,\omega_{2},\Delta d)$ is the hydrodynamic force due to the moving air caused by the oscillatory motion of the
plate. These forces induce a bending of the cantilever $F/k$ according to Hooke's law, where $k$ is the spring
constant of the cantilever. The output of the optical lever read-out $S$ is then
changed by $\Delta S=\gamma F / k$, where the sensitivity of the read-out is characterized by the
calibration factor $\gamma$. We will now develop the full form of this signal $\Delta S$.

Following elementary electrostatic arguments, one can show that the electrostatic
force between a plane and a sphere of radius $R$ is given by

\begin{equation}
F_E(V,d)=-\frac{\varepsilon_0 \pi R \left(V+V_0\right)^2}{d}, \label{elecforce}
\end{equation}

\noindent where $\varepsilon_0$ is the permittivity of vacuum, $V$ is the externally applied voltage, $V_0$ is the contact
potential difference between the two surfaces, and $d \ll R$ (i.~e.~within the \emph{proximity force approximation} (PFA)~\cite{parsegian}). To evaluate the total signal $\Delta S$, we substitute $V=V_{DC} + V_{AC} \cos \left( \omega_1 t \right)$ and incorporate the distance modulation $\Delta d \cos\left(\omega_2 t\right)$. We then approximate the photodetector signal with a first order Taylor expansion for small excursion $\Delta d \cos\left(\omega_2 t\right)$ around $d=d_{0}-d_{pz}$ (see Fig.~\ref{fig:schematic}~c):

\begin{equation}
\Delta S(t) \simeq S_0 + S_{\omega_1} \cos (\omega_1 t) + S_{2 \omega_1} \cos (2\omega_1
t)+S_{\omega_2}^I \cos (\omega_2 t) + S_{\omega_2}^Q \sin (\omega_2 t) +
S_{\mathrm{rem}}(t),\label{Stimeseries}
\end{equation}

\noindent where

\begin{equation}
S_0 = -\frac{\gamma \varepsilon_0 \pi R \left[ \left(V_0+V_{DC}\right)^2+V_{AC}^2 /
2\right]}{k\left(d_0-d_{pz}\right)}+\frac{\gamma}{k}F_C\left(d_0-d_{pz}\right),\label{S0}
\end{equation}

\begin{equation}
S_{\omega_1} = -\frac{2 \gamma \varepsilon_0 \pi R
\left(V_0+V_{DC}\right)V_{AC}}{k\left(d_0-d_{pz}\right)},\label{Somega1}
\end{equation}

\begin{equation}
S_{2\omega_1} = -\frac{\gamma \varepsilon_0 \pi R}{k
\left(d_0-d_{pz}\right)}\frac{V_{AC}^2}{2},\label{S2omega1}
\end{equation}

\begin{equation}
S_{\omega_2}^I=-\frac{\gamma \varepsilon_0 \pi R \left[ \left(V_0+V_{DC}\right)^2+V_{AC}^2 /
2\right]\Delta d}{k\left(d_0-d_{pz}\right)^2}-\frac{\gamma}{k}\left.\frac{\partial F_C}{\partial
d}\right|_{d_0-d_{pz}}\Delta d ,\label{Somega2}
\end{equation}

\begin{equation}
S_{\omega_2}^Q=\frac{\gamma}{k} F_H(d_0-d_{pz},\omega_{2},\Delta d),\label{Somega2Q}
\end{equation}

\noindent $F_C\left(d_0-d_{pz}\right)$ is the Casimir force at separation $d_0-d_{pz}$, $S_{\mathrm{rem}}(t)$ contains the cross terms at frequencies like $\omega_1 \pm
\omega_2$ and $2\omega_1 \pm \omega_2$ and the gradient of the hydrodynamic force at $2
\omega_2$, and we have neglected the effect of the cantilever deflection on the distance between the surfaces. Since the remaining terms in $S_{\mathrm{rem}}$ are located at different frequencies than our measurement signals, they do
not interfere with the lock-in measurements of $S_{\omega_1}$, $S_{2 \omega_1}$, $S_{\omega_2}^I$ and
$S_{\omega_2}^Q$. $S_{\mathrm{rem}}$ will thus be neglected in the rest of the paper.

Eq.~\ref{Stimeseries} is only valid if the force sensor can follow the modulations of the
force without picking up phase delays. It is thus convenient to operate in the quasi-static regime, which also ensures that the amplitude response of our cantilever at the various measurement frequencies does not vary. For these reasons, we set $\omega_1 / 2 \pi = 72.2$~Hz and $\omega_2 / 2 \pi = 119$~Hz, which are
both much lower than the resonance frequency of the force sensor (1.9~kHz, quality factor $\simeq
75$ in air). Furthermore, we have not included the elastic component of the hydrodynamic interaction in $S_{\omega_2}^I$. According to \cite{Maali:2008p96}, the compression effect is small as long as $\sigma_{sphere} = \frac{4 \eta \omega_2 R}{p d} <1$, where $\eta$ is the viscosity of air and $p$ is the air pressure. In our experiment, $\sigma_{sphere} \leq 10^{-3}$, so the elastic component can be neglected and we only have to consider the dissipative part of $F_H(d,\omega_{2})$. Since a dissipative effect depends on velocity $v=\partial d /
\partial t = \omega_2 \Delta d \sin(\omega_2 t)$, it will manifest itself as a cantilever oscillation at
$\omega_2$ with a corresponding detector signal $S_{\omega_2}^Q$ that is 90~degrees rotated with
respect to $S_{\omega_2}^I$.

\subsection{Electrostatic calibration}
The first task of the electrostatic calibration procedure is to compensate for the presence of the contact potential difference $V_0$ between the two interacting surfaces. Since $S_{\omega_1}$
is proportional to $V_0+V_{DC}$ (see Eq.~\ref{Somega1}), we can create a negative feedback
loop in which a lock-in amplifier at $\omega_1$ generates $V_{DC}$ in such a way that
$S_{\omega_1}$ vanishes, i.e.~$V_{DC}=-V_0$ \cite{deMan:2009p104,deMan:2010p1654}. The stability of this feedback loop is guaranteed by a single large time constant. In the current experiment, the systematic error in the compensation voltage is negligible ($|V_0+V_{DC}|<50~\mu$V), and the statistical error is $\simeq 1$~mV. This feedback scheme is similar to Kelvin probe force microscopy \cite{Nonnenmacher:1991p22}, and allows one to measure $V_0$ at all sphere-plane separations. Even more, the automatic compensation of $V_0$ leads to the zeroing of the $(V_0+V_{DC})$ terms in Eqs.~\ref{S0} and~\ref{Somega2}, greatly simplifying
the measurement scheme.

The periodic component of $\Delta S$ at $2 \omega_1$, $S_{2\omega_1}$, measured with a second lock-in amplifier (the calibration lock-in in Fig.~\ref{fig:schematic}~b), is used to calibrate the force sensitivity and to find the initial separation between the surfaces $d_{0}$. We define

\begin{equation}
\alpha = \frac{\gamma \varepsilon_0 \pi R}{k \left(d_0-d_{pz}\right)}. \label{alphadefinition}
\end{equation}

\noindent According to Eq.~\ref{S2omega1}, $\alpha$ can be experimentally obtained from $\alpha=2 \left|S_{2 \omega_1}\right| / V_{AC}^2$. In this way, we have essentially performed an AC measurement of the curvature of the electrostatic parabola, instead of using multiple DC measurements with different applied voltages~\cite{Iannuzzi:2004p117}. We measure $\alpha$ as a function of $d_{pz}$ by
varying the extension of the capacitive feedback controlled piezoelectric transducer (see
Fig.~\ref{fig:schematic}c) in discrete steps. We then fit Eq.~\ref{alphadefinition} to these $\alpha$ data. This procedure allows us to calibrate the separation at the start
of the measurement $d_0$ and the force sensitivity $\kappa = \gamma \varepsilon_0 \pi R / k$ for
each measurement run. We then use the estimate of $d_0$ to adjust the initial value of $d_{pz}$ of the next measurement run in order to have all runs start at the same separation. To avoid large electrostatic forces at small separations, we reduce $V_{AC}$
as the surfaces approach such that $S_{2 \omega_1}$ stays nearly constant at a value that corresponds to a
root-mean-square electrostatic force of $\simeq 50$~pN~\cite{deMan:2009p104}.

\subsection{Casimir force measurement}
We use a third lock-in amplifier (the measurement lock-in in Fig.~\ref{fig:schematic}~b), locked at $\omega_2$, to measure the Casimir force. The phase of
this lock-in amplifier is aligned to the actual motion of the plate by examining the signal from a dedicated fiber optic
interferometer (not shown in Fig.~\ref{fig:schematic}). The same interferometer is used to
calibrate the amplitude $\Delta d$ of the separation modulation. We
see from Eq.~\ref{Somega2} that the in-phase component $S_{\omega_2}^I$ contains both an
electrostatic contribution and the gradient of the Casimir force $F_C$ at the current
separation. Since $V_0+V_{DC}=0$ by the $V_0$ feedback circuit, Eq.~\ref{Somega2}
simplifies to

\begin{equation}
S_{\omega_2}^I=-\frac{\gamma \varepsilon_0 \pi
R}{k\left(d_0-d_{pz}\right)^2}\frac{V_{AC}^2}{2}\Delta d-\frac{\gamma}{k}\left.\frac{\partial
F_C}{\partial d}\right|_{d_0-d_{pz}}\Delta d.\label{Somega2simplified}
\end{equation}

\noindent Combining Eqs.~\ref{S2omega1} and~\ref{Somega2simplified}, one obtains

\begin{equation}
S_{\omega_2}^I=\frac{S_{2\omega_1}}{d_0-d_{pz}}\Delta d-\frac{\gamma}{k}\left.\frac{\partial
F_C}{\partial d}\right|_{d_0-d_{pz}}\Delta d.\label{Somega2final}
\end{equation}

\noindent Since the absolute separations $d_0-d_{pz}$ and $S_{2\omega_1}$ are known from the
simultaneous electrostatic calibration (and $\Delta d$ is calibrated too), one can calculate the value of
the first term of Eq.~\ref{Somega2final}. Using the force sensitivity $\kappa=\gamma
\varepsilon_0 \pi R / k$ obtained from the calibration, we can finally get the Casimir force gradient:
\begin{equation}
\frac{1}{R}\frac{\partial F_{C}}{\partial d} = \frac{\varepsilon_{0} \pi}{\kappa} \left( \frac{S_{2 \omega_{1}}}{d_{0}-d_{pz}} - \frac{S_{\omega_2}^I}{\Delta d} \right).
\label{eq:1overRdFdd}
\end{equation}
It is interesting to note that we obtain the Casimir force gradient divided by the sphere radius $R$, because we have calibrated the instrument with the electrostatic force which scales linearly in $R$ (see Eq.~\ref{elecforce}). However, within the PFA, the gradient of the force between a sphere and a plate relates directly to the pressure between two parallel plates $P_{pp}$ as long as $d \ll R$:
\begin{equation}
\frac{1}{R}\frac{\partial F_C}{\partial d}=2 \pi P_{pp}(d),
\label{eqpfa}
\end{equation}
where $P_{pp} (d)$ can be calculated with the Lifshitz theory~\cite{lifshitz} and depends only on the dielectric properties of the interacting surfaces. Therefore, we can directly compare our $1/R \ \partial F_{C}/\partial d$ data to theory, without any need to know the precise radius of the sphere.

Furthermore, by using a quadrature lock-in amplifier at $\omega_{2}$, we can obtain $S_{\omega_2}^Q$ together with $S_{\omega_2}^I$. We can thus measure the hydrodynamic interaction between the sphere
and the plate simultaneously with, but independently from, the Casimir force gradient.

\subsection{Determination of deflection sensitivity and cantilever spring constant}
So far, we have neglected the bending of the cantilever in the assessment of the distance between the sphere and plate surfaces. This is valid as long as the forces are relatively weak and the spring constant of the cantilever is relatively high. Of course, the nominal spring constant of the cantilever is supplied by the manufacturer, but the addition of a glued sphere and metal coating influence the stiffness. Therefore, we have developed a technique to measure the spring constant with the electrostatic force. Furthermore, this method also allows us to extract the deflection sensitivity $\gamma$ of the optical lever readout; we can then convert photodetector signal $\Delta S$ into cantilever deflection $F/k$. This technique might be useful for AFM force measurements in general.

To obtain the cantilever spring constant and the deflection sensitivity, we apply a relatively large $V_{AC}$ between the sphere and the plate. Exactly like described above in the electrostatic calibration section, we keep the electrostatic force at $2 \omega_{1}$ constant, but now at roughly 2~nN RMS instead of 50~pN RMS, by reducing $V_{AC}$ while increasing the piezoelectric transducer extension $d_{pz}$ in discrete steps. This strong force will reduce the sphere plate distance, and we therefore have to solve the following implicit equation for the electrostatic force

\begin{equation}
F_E=-\frac{\varepsilon_0 \pi RV^2}{d_{0}-d_{pz}+F_{E}/k} \label{eq:elecforcewithbending}
\end{equation}
where, since we have already dealt with the contact potential difference $V_{0}$ with the feedback circuit, $V$ just refers to the AC component of the applied voltage. For the sake of simplicity, we have omitted the piezo modulation at $\omega_{2}$ from this derivation, as it does not affect the results. Eq.~\ref{eq:elecforcewithbending} has two solutions for $F_{E}$, and the physically correct one reads
\begin{equation}
F_{E}=-\frac{1}{2} \left[ k (d_{0}-d_{pz}) - \sqrt{k^{2} (d_{0}-d_{pz})^{2} - 4 k \varepsilon_{0} \pi R V^{2}} \right].
\end{equation}
If we Taylor expand this expression for small cantilever deflection (which means small force and small applied voltage $V$), and use $\Delta S=\gamma F/k$, we obtain
\begin{equation}
\Delta S = -\frac{\gamma \epsilon_{0} \pi R V^{2}}{k (d_{0}-d_{pz})} - \frac{\gamma \epsilon_{0}^{2} \pi^{2} R^{2} V^{4}}{k^{2} (d_{0}-d_{pz})^{3}} + O(V^{6}).
\end{equation}
Substituting $V=V_{AC} \cos (\omega_{1} t)$ and neglecting the higher order terms yields a detector signal
\begin{equation}
\Delta S(t) \simeq S_0 + S_{2 \omega_1} \cos (2 \omega_1 t) + S_{4 \omega_1} \cos (4\omega_1 t),
\end{equation}
where $S_{0}$ is the DC component and the amplitudes of the two AC components are given by
\begin{equation}
S_{2 \omega_1} = -\frac{\gamma \epsilon_{0} \pi  R V_{AC}^{2}}{2 k (d_{0}-d_{pz})} - \frac{\gamma \epsilon_{0}^{2} \pi^{2} R^{2} V_{AC}^{4}}{2 k^{2} (d_{0}-d_{pz})^{3}}\label{eq:s2omega1withbigvoltage}
\end{equation}
and
\begin{equation}
S_{4 \omega_1} = -\frac{\gamma \epsilon_{0}^{2} \pi^{2} R^{2} V_{AC}^{4}}{8 k^{2} (d_{0}-d_{pz})^3}.
\end{equation}
$S_{2\omega_{1}}$ is already measured by our electrostatic calibration lock-in amplifier, and we simply add another lock-in amplifier locked at $4\omega_{1}$ to detect $S_{4\omega_{1}}$.

The second term in Eq.~\ref{eq:s2omega1withbigvoltage} is much smaller than the first term and can be neglected. We then find that
\begin{equation}
V_{AC} \sqrt{\frac{S_{2 \omega_{1}}}{S_{4 \omega_{1}}}} = 2 \sqrt{\frac{k}{\epsilon_{0} \pi R}}(d_{0}-d_{pz}),
\label{eq:4ffit}
\end{equation}
which means that we can obtain $k/R$ by fitting Eq.~\ref{eq:4ffit} to data of $V_{AC} \sqrt{S_{2\omega_{1}}/S_{4\omega_{1}}}$ as a function of relative piezo displacement $d_{pz}$. Apart from the resulting knowledge on the cantilever spring constant (the sphere radius is roughly known), we also obtain the deflection sensitivity $\gamma$ by combining the value of $k/R$ with the one of $\kappa = \gamma \varepsilon_0 \pi R / k$ determined by the analysis of the simultaneously acquired $\alpha$ data (as described in the electrostatic calibration procedure).

\section{Results and Discussion}

We have divided our experimental results into two parts. In the first part, we describe the precision and stability of the electrostatic calibration procedure and comment on the reproducibility of the Casimir force gradient detection in a set of 580 measurement runs between two gold coated surfaces. Also, we present a single dataset obtained with a large electrostatic force between the sphere and the plate that allows us to check the validity of Eq.~\ref{Somega2} (and Eq.~\ref{eq:1overRdFdd} as well), and to obtain the spring constant of our cantilever and the deflection sensitivity of the optical lever readout. In the second half, we combine the Casimir force measurements between the two gold surfaces with measurements between a gold surface and a surface coated with ITO (In$_2$O$_3$:Sn), as presented in~\cite{deMan:2009p99}, adding details that, for the sake of brevity, were previously omitted. Furthermore, we obtain the hydrodynamic forces for both measurement sets and compare the results.

\subsection{General performance}

We will now analyze the 580~measurement runs between two gold surfaces obtained during nearly 72~hours of continuous data acquisition. In this experiment, the separation between the surfaces is varied in discrete steps with the feedback controlled piezoelectric transducer, and a typical measurement
run consists of $\simeq 50$ $d_{pz}$ set points in the measurement range $50 < d < 1100$~nm. The lock-in measurements are obtained with 24~dB roll-off low-pass filter settings with 1~s RC time. The waiting time for every value of $d_{pz}$ is 8~s, and a complete run takes roughly 7 minutes. The $S_{2\omega_1}$ set point corresponds to a cantilever movement of approximately 50~pm~RMS. The distance modulation is set to $\Delta d = 3.85 \pm 0.08$~nm, and the in-phase and out-of-phase cantilever responses at $\omega_2$ are $<80$~pm RMS during the entire experiment.
All force
measurements are performed in air at atmospheric pressure, temperature 300~K, and 29\% relative
humidity.

\begin{figure}[h!]
\includegraphics[width=10cm]{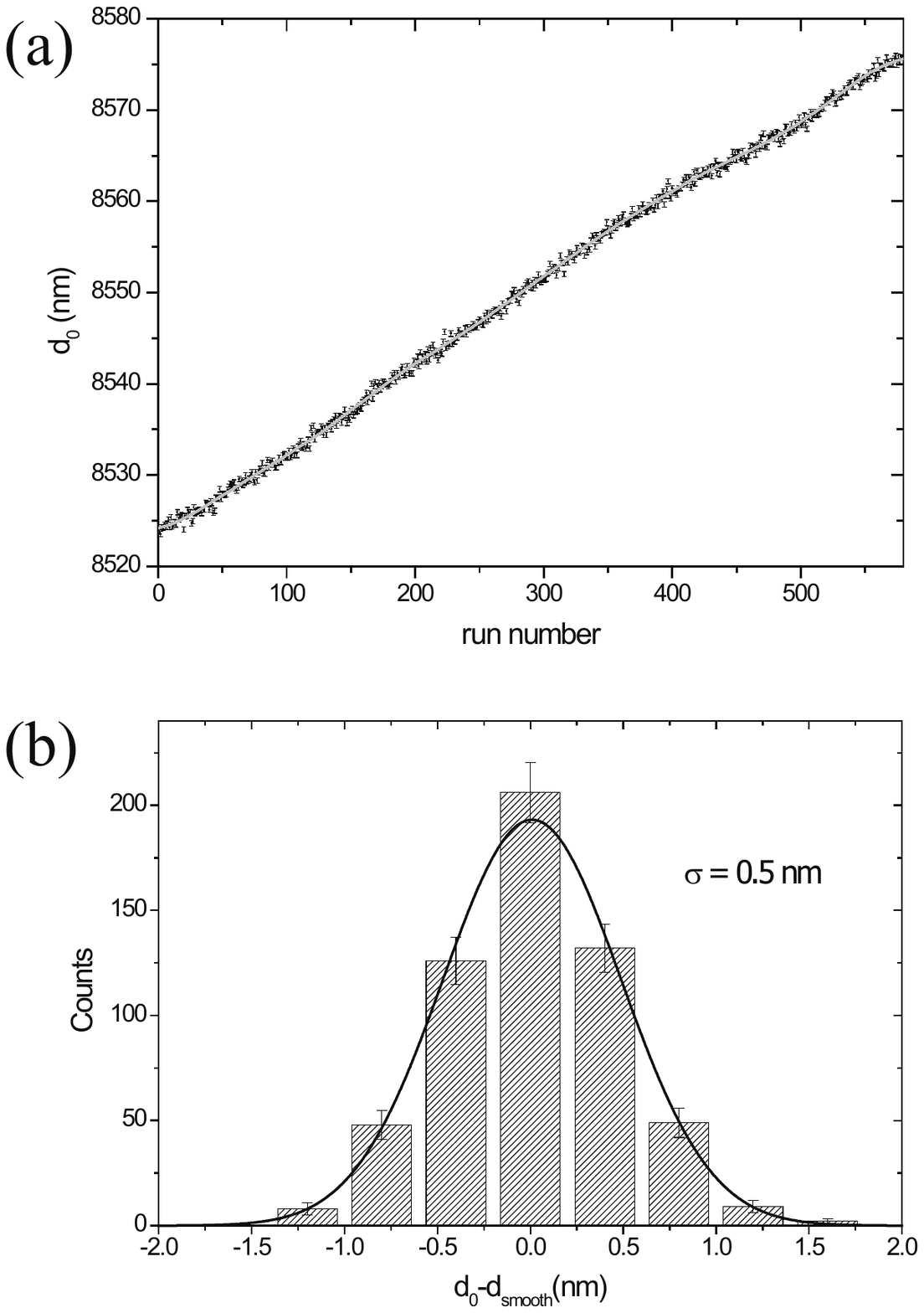}
\caption{Mechanical stability of the experimental setup. (a) Mechanical drift in the initial separation
$d_0$ as a function of run number for all 580 Au-Au measurement runs. The error bars
are determined by propagation of the error on $\alpha$ into the estimate of $d_0$ by the fit with
Eq.~\ref{alphadefinition}. The grey line represents a trend line that accounts for the slow thermal drift of the setup. (b) Histogram of the differences between the measured $d_{0}$ values and the grey line of a. The line represents the best Gaussian fit, resulting in a 0.5~nm standard deviation.}
\label{fig:eleccalibration}
\end{figure}

\begin{figure}[h!]
\includegraphics[width=10cm]{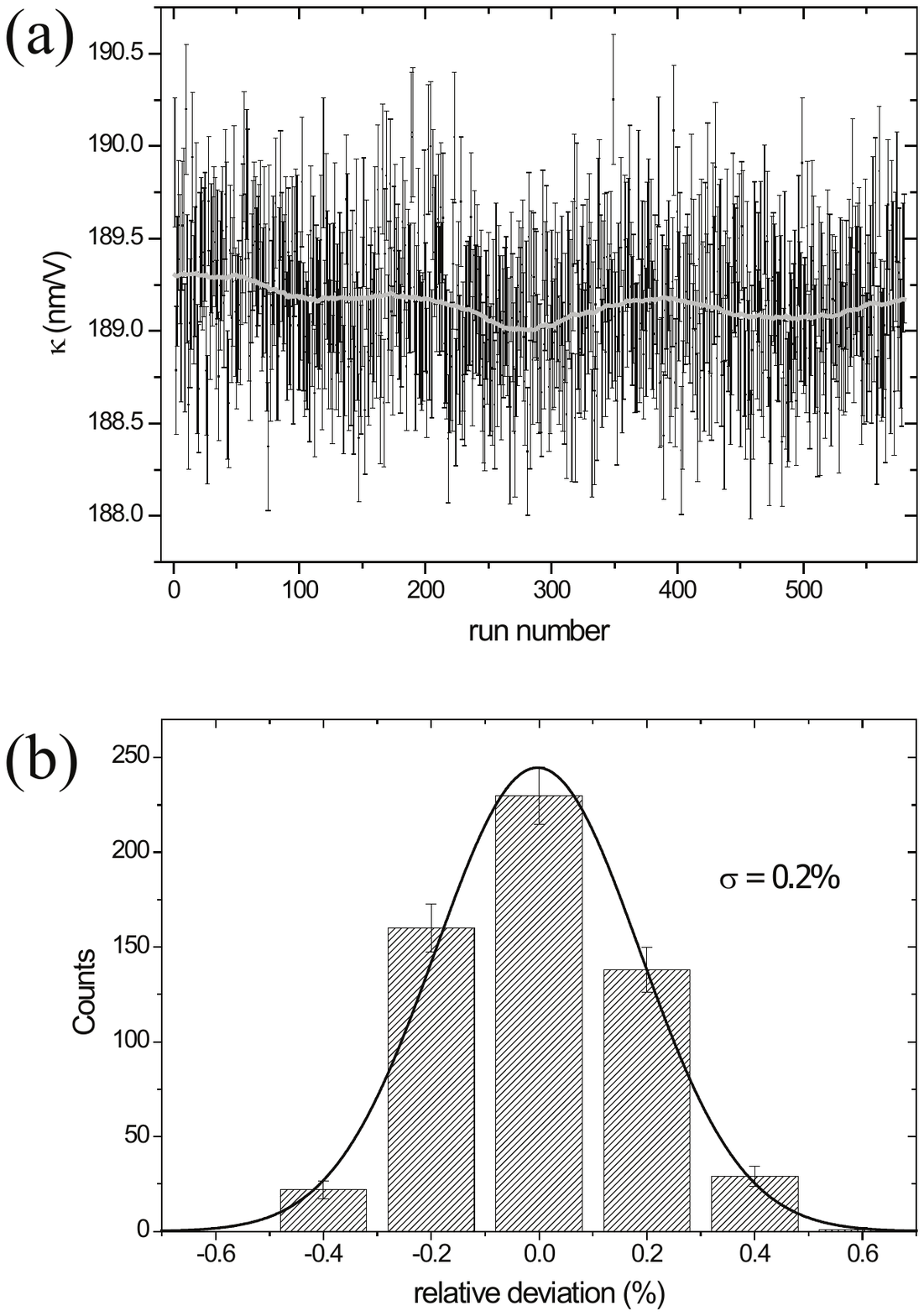}
\caption{Stability of the electrostatic calibration. (a) All 580 obtained values for the force calibration constant $\kappa$ as a function of run number. The error bars are calculated by propagating the error on $\alpha$. The grey line is a smooth trend line that accounts for slow variations. (b) Histogram of the relative deviations of the $\kappa$ data from the trend line in c. The Gaussian fit has a standard deviation of 0.2\%.}
\label{fig:eleccalibration2}
\end{figure}

Concerning the electrostatic calibration, we have to fit our $\alpha$ data with Eq.~\ref{alphadefinition} to obtain the initial separation $d_0$ and the force sensitivity $\kappa$. Due to the fact that we hold $S_{2\omega_1}$ constant by reducing $V_{AC}$, the relative statistical error in $\alpha$ is constant (see reference \cite{deMan:2009p104}) and was measured to be $\simeq 0.7\%$. We have verified that $\alpha$ follows Eq.~\ref{alphadefinition}, as suggested in \cite{Kim:2008p55} and discussed in \cite{deMan:2009p104}. In Fig.~\ref{fig:eleccalibration} and Fig.~\ref{fig:eleccalibration2}, we present the fitted values for $d_0$ and $\kappa$ and analyze their stability in time. Fig.~\ref{fig:eleccalibration}~a shows all the values of $d_0$ for the 580 runs with error bars as propagated from the error on $\alpha$. The grey line represents the smooth thermal expansion of the setup, and is estimated by smoothing the data with a 100~point moving window second order Savitsky-Golay filter. The total mechanical drift of our setup is 52~nm in 72~hours, which is less than 1~nm per hour and less than 0.1~nm per measurement run. Clearly, we can neglect the mechanical drift in our assessment of separation between the surfaces in a single run. In Fig.~\ref{fig:eleccalibration}~b, we plot a histogram of the difference between the $d_0$ data and the smoothed grey line of Fig.~\ref{fig:eleccalibration}~a. These differences are clearly normally distributed with a standard deviation of 0.5~nm. Therefore, in these experimental runs, we could determine the separation between the sphere and plate surfaces with 0.5~nm precision. This estimate of the precision in the measurement of $d_0$ is insensitive to the precise form and size of the smoothing window.

Fig.~\ref{fig:eleccalibration2}~a shows all the values of the force sensitivity $\kappa$ that we obtained from the fit to our electrostatic calibration data. The error bars are calculated by propagating the errors on $\alpha$. The grey line is a smoothed trend line that represents slow variations in $\kappa$ over time, obtained by smoothing the data with a 200~point moving window second order Savitsky-Golay filter. There is clearly no long-term drift in the force sensitivity, which shows that our setup is very stable. In Fig.~\ref{fig:eleccalibration2}~b, we plot a histogram of the relative deviations between our $\kappa$ data and the smooth grey line of Fig.~\ref{fig:eleccalibration2}~a. These deviations are normally distributed with a standard deviation of 0.2\%, and are insensitive to the specifics of the smoothing window. We have thus determined the force sensitivity of the setup for every single measurement run with a precision of 0.2\%.

\begin{figure}[h!]
\includegraphics[width=13cm]{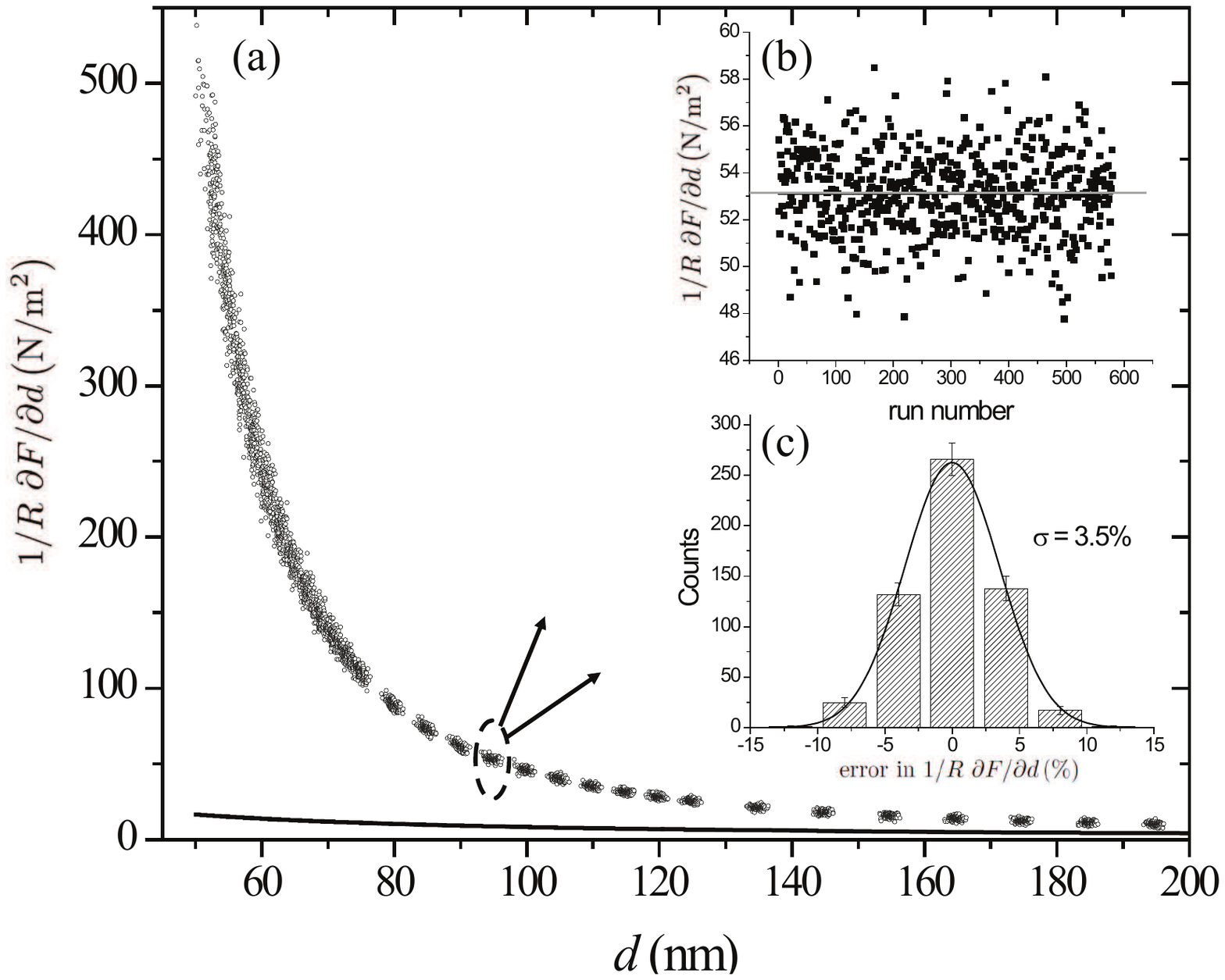}
\caption{(a) The data points represent measurements of the total force gradient as a function of separation between the sphere and plate surfaces for 150 measurement runs out of a total of 580 runs. The line shows the electrostatic force gradient associated to the simultaneous calibration procedure. (b) Plot of all 580 force gradient measurements obtained for $d \simeq 95$~nm as a function of time. The grey line represents the average force gradient. (c) Histogram of all the relative deviations between the single force gradient measurements around 95~nm and the average force gradient. The Gaussian fit has a standard deviation of 3.5\%.} \label{fig:forceresults}
\end{figure}

In Fig.~\ref{fig:forceresults} we present measurements of the total force gradient
\begin{equation}
\frac{1}{R}\frac{\partial F}{\partial d} = -\frac{\varepsilon_{0} \pi}{\kappa} \frac{S_{\omega_2}^I}{\Delta d}
\label{eq:totalforcegradientmeasurement}
\end{equation}
as a function of the non-modulated separation $d=d_{0}-d_{pz}$ (see Fig.~\ref{fig:schematic}~c). This force gradient should, according to Eq.~\ref{eq:1overRdFdd}, obey
\begin{equation}
\frac{1}{R}\frac{\partial F}{\partial d} = \frac{1}{R}\frac{\partial F_C}{\partial d} + \frac{1}{R}\frac{\partial F_E}{\partial d}
\end{equation}
with
\begin{equation}
\frac{1}{R}\frac{\partial F_E}{\partial d} = -\frac{\varepsilon_{0} \pi}{\kappa} \frac{S_{2 \omega_{1}}}{d_{0}-d_{pz}}.
\label{eq:elecforcegradient}
\end{equation}
In Fig.~\ref{fig:forceresults}~a, the data points represent the $-\varepsilon_{0} \pi S_{\omega_2}^I / (\kappa \Delta d)$ data points and the solid line shows the electrostatic force gradient as obtained with Eq.~\ref{eq:elecforcegradient} from the $S_{2\omega_1}$ values of the simultaneous electrostatic calibration procedure. For clarity, we have shown only 150 measurement runs out of the total 580. It is clear that in the distance range that our setup is sensitive for the Casimir force, the electrostatic force gradient caused by the simultaneous electrostatic calibration is small compared to the Casimir force gradient. To assess the stability of our force gradient measurement, we have plotted all 580 $1/R \ \partial F/\partial d$ data points gathered around 95~nm from our 580 measurement runs in Fig.~\ref{fig:forceresults}~b. Our data do not show any drift in time, which means that the setup is stable. The grey line represents the average of the data points. In Fig.~\ref{fig:forceresults}~c, we plot a histogram of all the relative deviations of the data with respect to the average. These deviations are normally distributed with a standard deviation of 3.5\%, which corresponds to a standard deviation in the measurement of the force gradient of 1.85~N/m$^2$. This value represents an overestimate of the noise though, because the data are obtained at slightly different separations; the exact position of a data-point depends on the estimate of $d_0$ coming from the previous measurement run. Since the error in the determination of $d_0$ is 0.5~nm (see Fig.~\ref{fig:eleccalibration}~b), the data are horizontally scattered with a standard deviation of 0.5~nm.  For $d \simeq 95$~nm, the local slope of the data in Fig.~\ref{fig:forceresults}~a is approximately 1.3~Nm$^{-2}$/nm, which translates this scatter in $d$ into a force gradient scatter of 0.65~N/m$^2$. Therefore, the actual precision in a single force gradient data point around 95~nm is 1.75~N/m$^2$, if we assume that both the force gradient noise and the scatter in separation are uncorrelated.

From the electrostatic calibration results, we could have also estimated the noise in the force gradient measurement. In fact, the noise in $S_{2\omega_{1}}$ is 30~$\mu$V~RMS with a 1~s RC time. The force gradient signal at $\omega_{2}$ is located at a comparable frequency, therefore the noise will be quite the same. If we substitute our measured values of $\kappa$ and $\Delta d$ into Eq.~\ref{eq:totalforcegradientmeasurement}, we see that we would have expected the noise in $1/R \ \partial F/\partial d$ to be 1.62~N/m$^2$. But we have not taken into account yet the 0.5~nm error in the separation that arises from the estimate of $d_{0}$. For $d \simeq 95$~nm, this results in an additional statistical error of 0.65~N/m$^2$ in the force gradient at this distance. The combined error, assuming the force gradient and distance errors are uncorrelated, is then 1.75~N/m$^2$ for $d \simeq 95$~nm, which agrees perfectly with the data of Fig.~\ref{fig:forceresults}.

\begin{figure}[!]
\includegraphics[width=13cm]{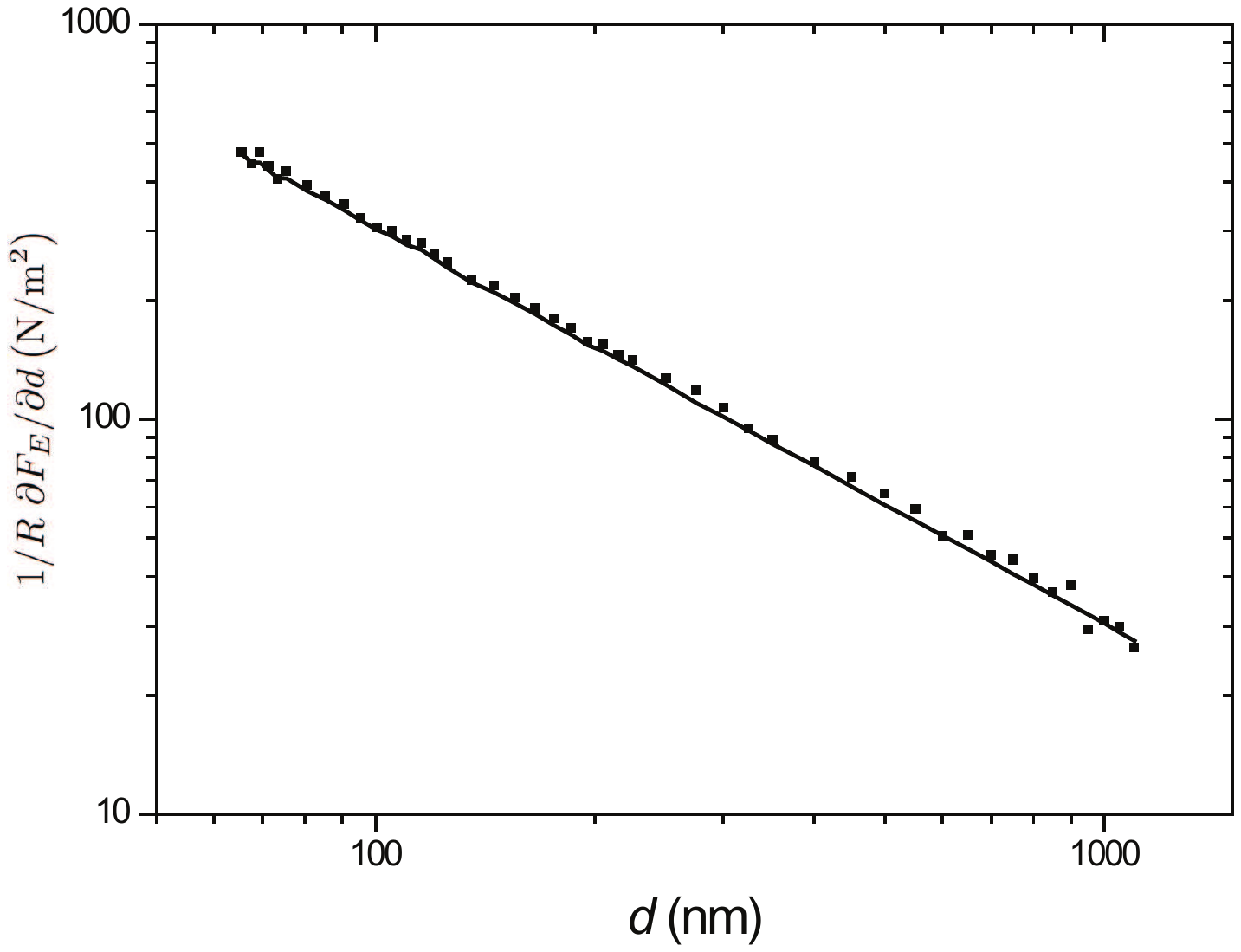}
\caption{
Plot of the electrostatic force gradient difference between a measurement run performed with a strong electrostatic interaction and a run performed with a weak electrostatic force. Data are plotted as a function of separation. The line corresponds to the electrostatic force gradient obtained from the calibration signal. See text for details.
}
\label{fig:checkgradient}
\end{figure}

Since our Casimir force gradient measurement consists of measuring the total force gradient and subtracting the electrostatic force gradient (see Eq.~\ref{eq:1overRdFdd}), it is interesting to investigate the accuracy in the assessment of $1/R \ \partial F_E/\partial d$. For that, we have gathered a new dataset with a relatively strong electrostatic interaction (high $V_{AC}$) between the sphere and the plate. When we combine the high $V_{AC}$ total force gradients with measurements obtained with low $V_{AC}$, we can get
\begin{equation}
\left.\frac{1}{R}\frac{\partial F}{\partial d}\right|_{V_{AC}>} - \left.\frac{1}{R}\frac{\partial F}{\partial d}\right|_{V_{AC}<} = \left.\frac{1}{R}\frac{\partial F_E}{\partial d}\right|_{V_{AC}>} - \left.\frac{1}{R}\frac{\partial F_E}{\partial d}\right|_{V_{AC}<}
\label{eq:elecfrocegradientdifference}
\end{equation}
because the Casimir force gradient is equal in both cases and drops out. Even more, any other systematic effects present in the force gradient measurement that do not depend on $V_{AC}$, like, for example, laser light that reflects from the planar sample and hits the photodetector, are also cancelled in this way. The right-hand-side of Eq.~\ref{eq:elecfrocegradientdifference} can be calculated with Eq.~\ref{eq:elecforcegradient}, and we can thus assess the validity of the latter and, consequently, of Eq.\ref{eq:1overRdFdd}. In Fig.~\ref{fig:checkgradient}, we have plotted the difference in total force gradients (obtained with Eq.~\ref{eq:totalforcegradientmeasurement}) as a function of distance. The solid line represents the difference in calculated electrostatic force gradients (Eq.~\ref{eq:elecforcegradient}), determined with the corresponding sets of $S_{2\omega_1}$ data. Although the agreement between the two electrostatic force gradients is good (there are no adjustable parameters), there exists a slight discrepancy between the two curves. The measured total force gradient difference is systematically about 3\% higher than the values calculated from $S_{2\omega_1}$. If this discrepancy means that there is a small error in the determination of the electrostatic force gradient, then the measurement of the Casimir force gradient is almost unaffected. For example, for all $d<120$~nm the electrostatic force gradient is always $<25$\% of the total force gradient, which results in an error of $<1$\% in the measurement of the Casimir force gradient. If, on the other hand, the mismatch is caused by the uncertainty in the determination of $\Delta d$ with the dedicated fiber optic interferometer, then our Casimir force gradients are affected by a 3\% systematic error. Nevertheless, this systematic error will not hamper the comparison between force gradient data obtained with different samples, as we always use the same $\Delta d$.

\begin{figure}[!]
\includegraphics[width=13cm]{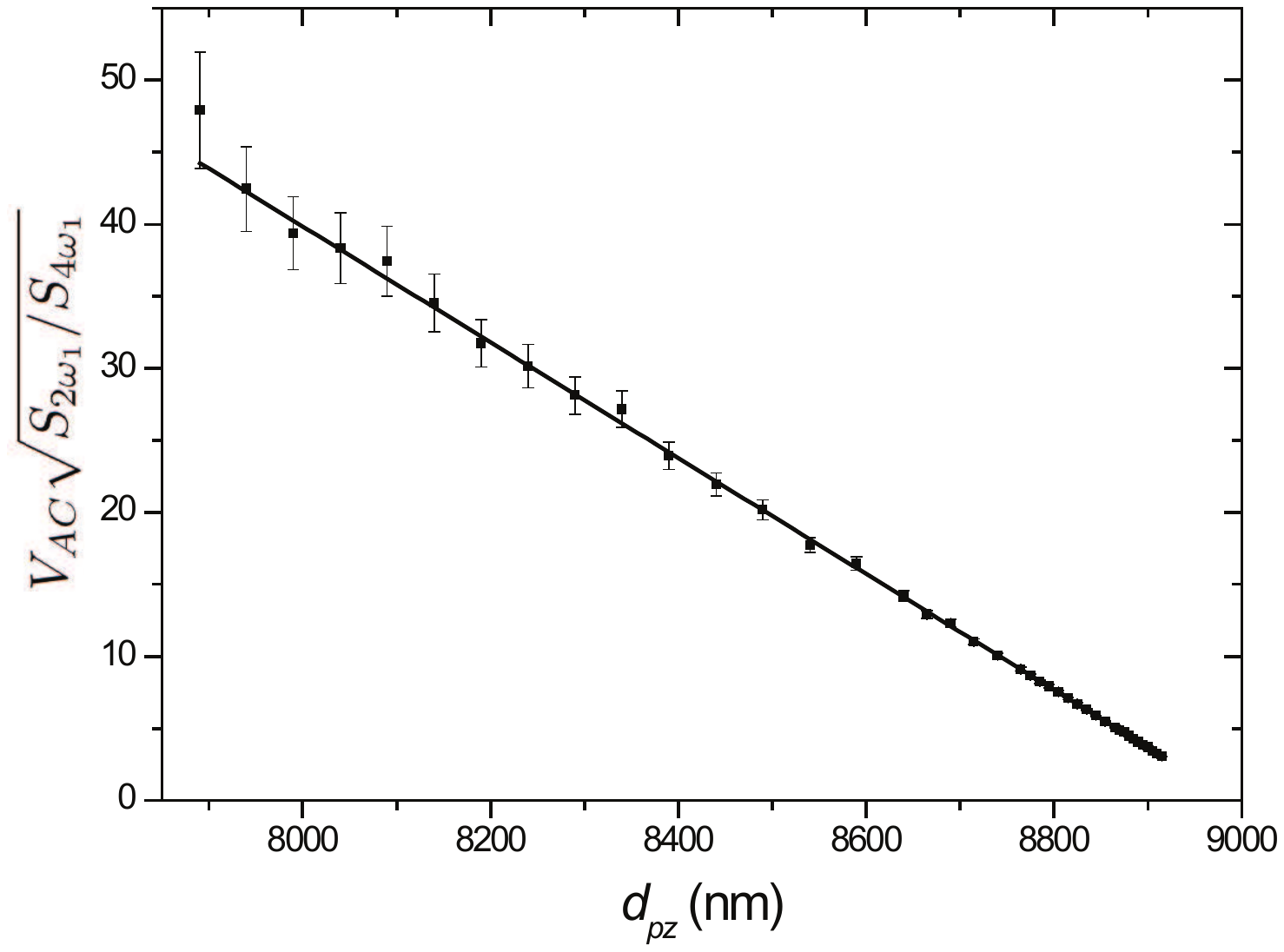}
\caption{Plot of $V_{AC} \sqrt{S_{2\omega_{1}}/S_{4\omega_{1}}}$ as a function of piezoelectric transducer extension. The line represents the best fit of the data with Eq.~\ref{eq:4ffit} (reduced $\chi^{2}=0.25$). The slope of the fit can be used to obtain the cantilever spring constant and the deflection sensitivity $\gamma$.
}
\label{fig:4f}
\end{figure}

To measure the deflection sensitivity of the readout and the spring constant of the cantilever, we follow the procedure outlined above. In essence, we apply a big potential difference between the sphere and the plate, record both the cantilever deflection signal at $2 \omega_{1}$ and $4 \omega_{1}$ as a function of relative piezo-electric transducer displacement $d_{pz}$, and fit Eq.~\ref{eq:4ffit} to those data. In Fig.~\ref{fig:4f}, we plot $V_{AC} \sqrt{S_{2\omega_{1}}/S_{4\omega_{1}}}$ as a function of $d_{pz}$ for such a single dataset. The straight line represents the best fit with Eq.~\ref{eq:4ffit} (reduced $\chi^{2}=0.25$). The error bars were determined by measuring the absolute error in $S_{2\omega_{1}}$ and assuming that the error in $S_{4\omega_{1}}$ is equal and independent from the error in $S_{2\omega_{1}}$. This is not entirely correct, because some sources of error, like for example fluctuations in $d$, will lead to correlated variations in $S_{2\omega_{1}}$ and $S_{4\omega_{1}}$. We have thus overestimated the error in $V_{AC} \sqrt{S_{2\omega_{1}}/S_{4\omega_{1}}}$, which leads to a reduced $\chi^{2}<1$. Anyhow, the slope of the data allows us to extract $k/R = (11.12 \pm 0.06) \ 10^{3} \ \mathrm{N}/\mathrm{m}^{2}$ (the uncertainty is obtained by setting reduced $\chi^{2}=1$). When we combine this value of $k/R$ with the simultaneously determined $\kappa = 191.3 \pm 0.2$~nm/V, we find that $\gamma = (7.64 \pm 0.04) \ 10^{7}$~V/m. With this value of $\gamma$, we can now establish that the $S_{2 \omega_{1}}$ set point we used for this dataset corresponds to a cantilever motion of 2~nm RMS at $2 \omega_{1}$. It is interesting to observe that this 2~nm modulation of the separation $d$ at $2 \omega_{1}$ gives rise to a measurable signal at $4 \omega_{1}$ even at 1~$\mu$m distance. Furthermore, if we use the approximately known sphere radius of 100~$\mu$m, we obtain the spring constant of our cantilever $k=1.1$~N/m. As the nominal spring constant before sphere attachment and gold coating was 0.9~N/m, the value we find with this electrostatic method is very reasonable.

With the deflection sensitivity calibrated, we can now assess the total cantilever bending and the precision in the measurements of the cantilever deflection. In the measurement runs presented in Figs.~\ref{fig:eleccalibration}, \ref{fig:eleccalibration2} and \ref{fig:forceresults}, we used an $S_{2\omega_{1}}$ set point of 4~mV RMS, which corresponds to a cantilever motion of 52~pm RMS. Therefore, the static bending of the cantilever due to the electrostatic calibration procedure is 74~pm (see Eqs.~\ref{S0} and~\ref{S2omega1}). Anyway, this static bending is constant during the measurement run and it is thus automatically taken into account in the estimate of $d_{0}$. The cantilever oscillations at $\omega_{2}$ caused by the total force gradient and the hydrodynamic interaction are $< 80$~pm RMS for these measurement runs, which means that the corresponding static bending is $<113$~pm. It is thus evident that we can safely neglect the static bending of the cantilever in our data analysis. Since the noise in $S_{2\omega_{1}}$ is 30~$\mu$V~RMS, the precision in the detection of the cantilever deflection is 400~fm~RMS with our 1~s RC time (24~dB low-pass filter). This means that our setup has an RMS sensitivity of 1~pm/$\sqrt{\mathrm{Hz}}$ at $2\omega_{1}/2\pi=144.4$~Hz.

\subsection{Halving the Casimir force}

We now present a comparison between two experiments performed with the same gold coated 100~$\mu$m radius sphere and two different plates. The first experiment is conducted with a polished sapphire substrate coated with a gold film similar to the one deposited on the sphere. The general performance of our setup was discussed above by analyzing this first experiment. The second experiment consists of 580~measurement runs in which the plate is replaced by a float glass substrate with a sputtered ITO thin film on top (PGO CEC010S, typically $8.5~\Omega/\Box$, or,
equivalently, $\rho = 1.6~10^{-4}~\Omega$cm). After purchase, this sample has been exposed to air
for more than two years before our measurements were performed.

\begin{figure}[!]
\includegraphics[width=13cm]{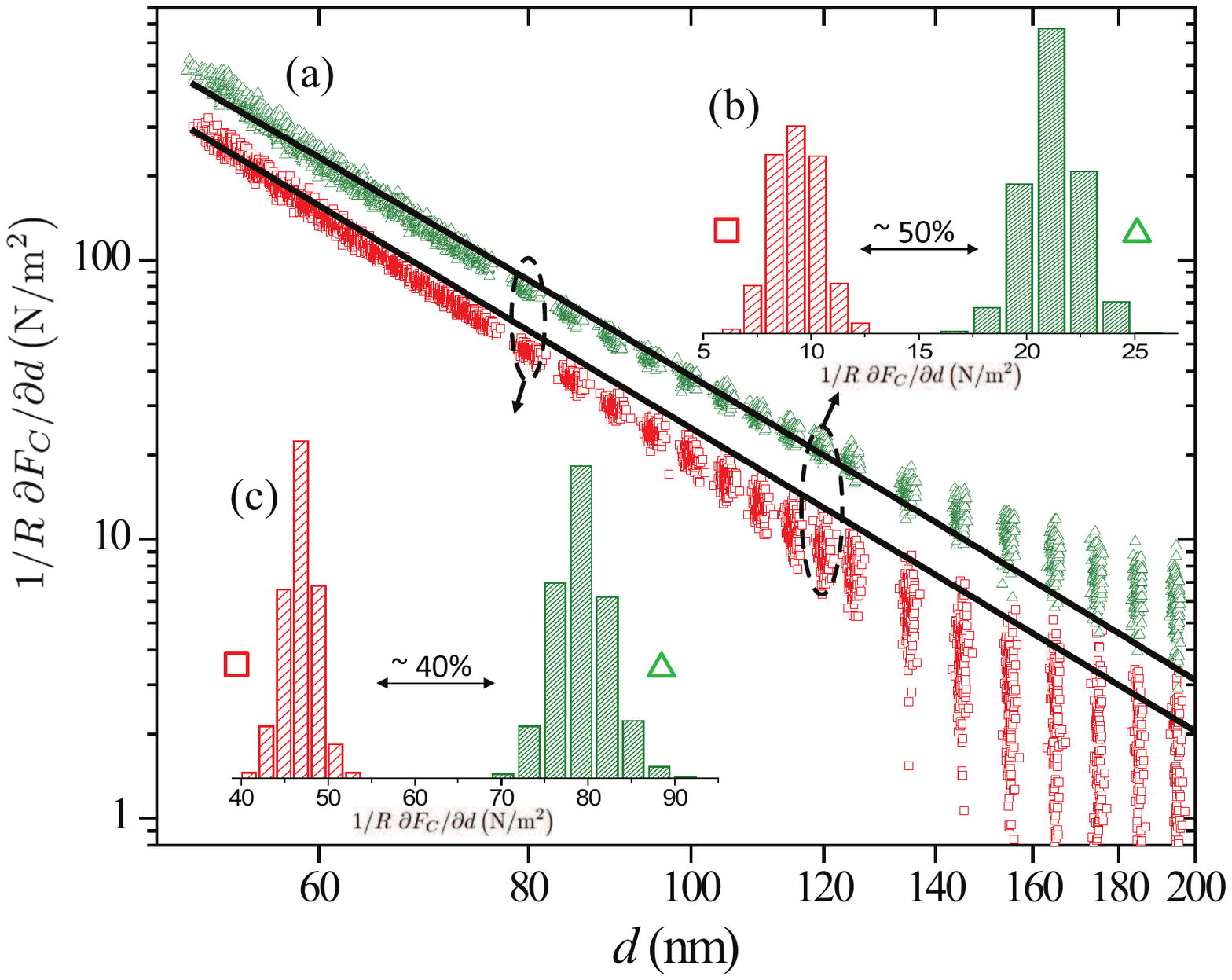}
\caption{(Color online) (a) Casimir force gradient as a function of
separation for the Au-Au (green triangles) and Au-ITO (red squares) interactions for
randomly chosen subsets of the data (150 out of 580 for both cases) plotted on
a double logarithmic scale, with the common electrostatic background subtracted from the data. The black
lines correspond to the calculated Casimir interactions. (b) Histogram of all 580 force
measurements for both the Au-Au and Au-ITO measurements at $d=120$~nm. The difference in Casimir
force gradient is $\simeq 50\%$ between the gold and ITO measurements. (c) Same as b, but
for $d=80$~nm. At this separation, the difference in the force gradient is $\simeq 40\%$.} \label{fig:ITO}
\end{figure}

Fig.~\ref{fig:ITO} shows the Casimir force gradient between the two pairs of surfaces (Au-Au in green triangles, Au-ITO in red squares)~\cite{deMan:2009p99}. In Fig.~\ref{fig:ITO}~a, we plot the force gradients as a function of separation on a double logarithmic scale for randomly chosen subsets of the data (150 out of 580 for both cases). Both datasets are obtained with the exact same settings for the electrostatic calibration and the force gradient measurement, and the Casimir force gradient is obtained from Eq.~\ref{eq:1overRdFdd}. The black lines indicate the theoretical force gradient, as will be explained below. Figs.~\ref{fig:ITO}~b and~c present two histograms of all 580~Casimir force gradient measurements for both Au-Au and Au-ITO at separations $d=120$~nm and $d=80$~nm, respectively. It is clear that the interaction strength with the ITO sample is considerably reduced with respect to the gold plate.

Note that our estimate of $d_0$, and thus $d$, relies on the simple form of Eq.~\ref{elecforce} and is only valid
for $d \ll R$ (PFA). This assumption is not entirely correct in the probed separation range \cite{deMan:2009p104}
and results in a systematic error in $d_0$ of about $1.4$~nm. Still, the corresponding
underestimate of the separation is equal for both the measurements with Au and ITO, and can thus be
neglected in the comparison of the two experiments.

Concerning the compensation voltage, we observed that $V_0$ varies approximately $1$~mV and $3$~mV
over the complete measurement range in the Au-Au and Au-ITO cases, respectively. These slight
variations of $V_0$ do not compromise the measurement of the Casimir force at the current level of
sensitivity. The value of $V_0$ drifts in time from $-106$ to $-103$~mV for Au-Au and from 72 to 50~mV for Au-ITO at $d=100$~nm. It is also important to note that, during the whole duration of the experiment, we
never observed any problem with electrostatic charging of the Au or ITO layers, which would have
most likely resulted in erratic behavior of $\alpha$ and/or $V_0$.

\begin{figure}[h!]
\includegraphics[width=13cm]{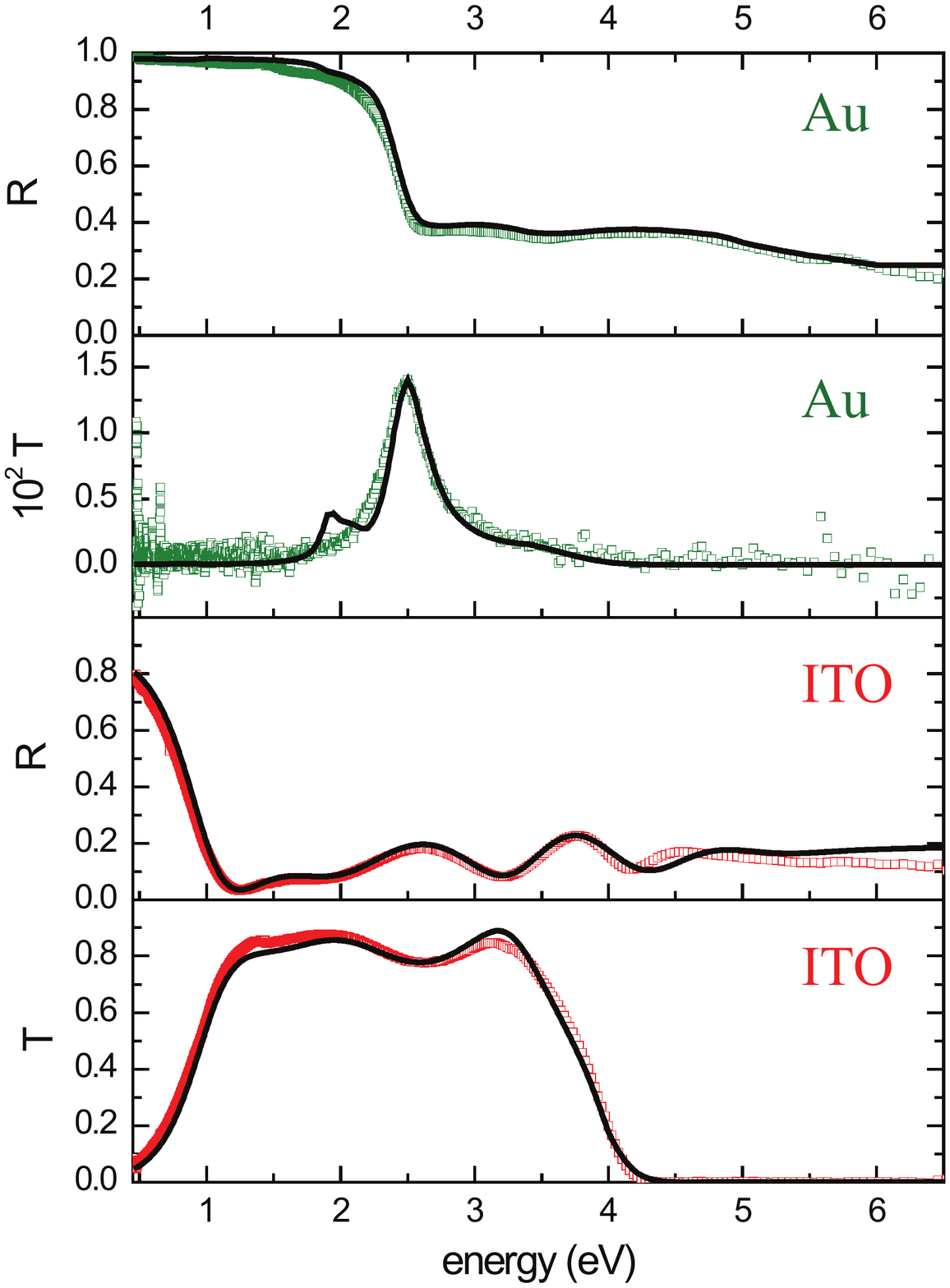}
\caption{(Color online) Measured reflection (R) and transmission (T)
spectra as a function of photon energy for the Au on sapphire (green data) and ITO on float
glass (red data) samples. The continuous black lines are calculations of the reflection and transmission
spectra expected for our samples (no fit parameters except ITO layer thickness), using handbook data for gold \cite{palik} and
a model from \cite{Fujiwara:2005p105} for ITO. The transmission spectra for gold are
zoomed in because the maximum transmission (around 2.5~eV) is only 1.4\%. The
calculation of the transmission spectrum of the ITO sample is quite sensitive to the
choice of dielectric properties of the float glass for photon energies above 4~eV. The black lines describe the measured data
reasonably enough to allow for calculations of the Casimir force.} \label{fig:opticalproperties}
\end{figure}

In order to compare the obtained Casimir force gradients with theoretical predictions, we have investigated the dielectric properties of our surfaces. In Fig.~\ref{fig:opticalproperties}, we show the reflection and transmission spectra of the two
plates, measured from the thin film side, in the frequency range from 0.5 to 6.5~eV. The black
continuous lines represent the reflection and transmission spectra calculated from the literature. For
Au, we used the values reported in \cite{palik}. The imaginary part of the dielectric
function of ITO is constructed from a sum of Drude and Tauc-Lorentz models with the parameters from
\cite{Fujiwara:2005p105}. The real part of the dielectric function is calculated with direct Kramers-Kronig
integration. The thickness of the ITO thin film is fitted by examining the interference fringes in the reflection and transmission spectra (taking into account the refractive index of the material) and turned out to be 190~nm, which is close to the typical thickness reported by the manufacturer (180~nm). The agreement between the spectroscopic measurements and these literature values is
reasonable in the probed energy range. We want to stress that for Au there are no adjustable parameters whatsoever in Fig.~\ref{fig:opticalproperties}, and that for ITO only the thickness was fitted. These results allow us to estimate the Casimir force expected in the
two cases (Au and ITO) and compare the calculation with our measurements (see Eq.~\ref{eqpfa}).

The theoretical Casimir interaction is calculated with the Lifshitz equation using the dielectric properties of our surfaces. For Au, we have extrapolated the data of~\cite{palik} with a Drude model ($\omega_p=9.0$~eV and $1/\tau=0.035$~eV from~\cite{Bordag:2001p118}). For ITO, we used the model from \cite{Fujiwara:2005p105} for all frequencies. The computed force gradient is plotted as the black lines in Fig.~\ref{fig:ITO}. The agreement between data and theory is reasonable, although we do seem to obtain different powers for data and theory. At small separation, the experimental curves are bending upwards, which is a sign of surface roughness effects~\cite{Maradudin:1980p1905}; the theoretical curves were calculated for perfectly smooth surfaces. Furthermore, the Au-Au data tend to give rise to a stronger force at large distance compared to theory, which is most likely caused by an artifact common to many AFM force measurements: the laser light is reflected from the planar sample into the photodetector giving rise to a background signal. In the Casimir force gradient method presented here, this artifact results, in first order, to an offset in the data; this explains the upwards trend of the data for large $d$. Although the precise distance dependence and strength of this artifact is unknown, we estimate from the force gradient data at large separation ($d>500$~nm) that the associated systematic error is certainly $<2$~N/m$^2$. In the case of the Au-ITO measurements, such a background signal is a lot smaller because ITO does not reflect well the laser light (see Fig.~\ref{fig:opticalproperties} at $\omega = 1.9$~eV). To explain the mismatch, it is therefore more likely that the model we used for the dielectric properties~\cite{Fujiwara:2005p105} is too metallic at low energy and that, consequently, the calculated Casimir interaction is too strong especially at large $d$.

\begin{figure}[!]
\includegraphics[width=13cm]{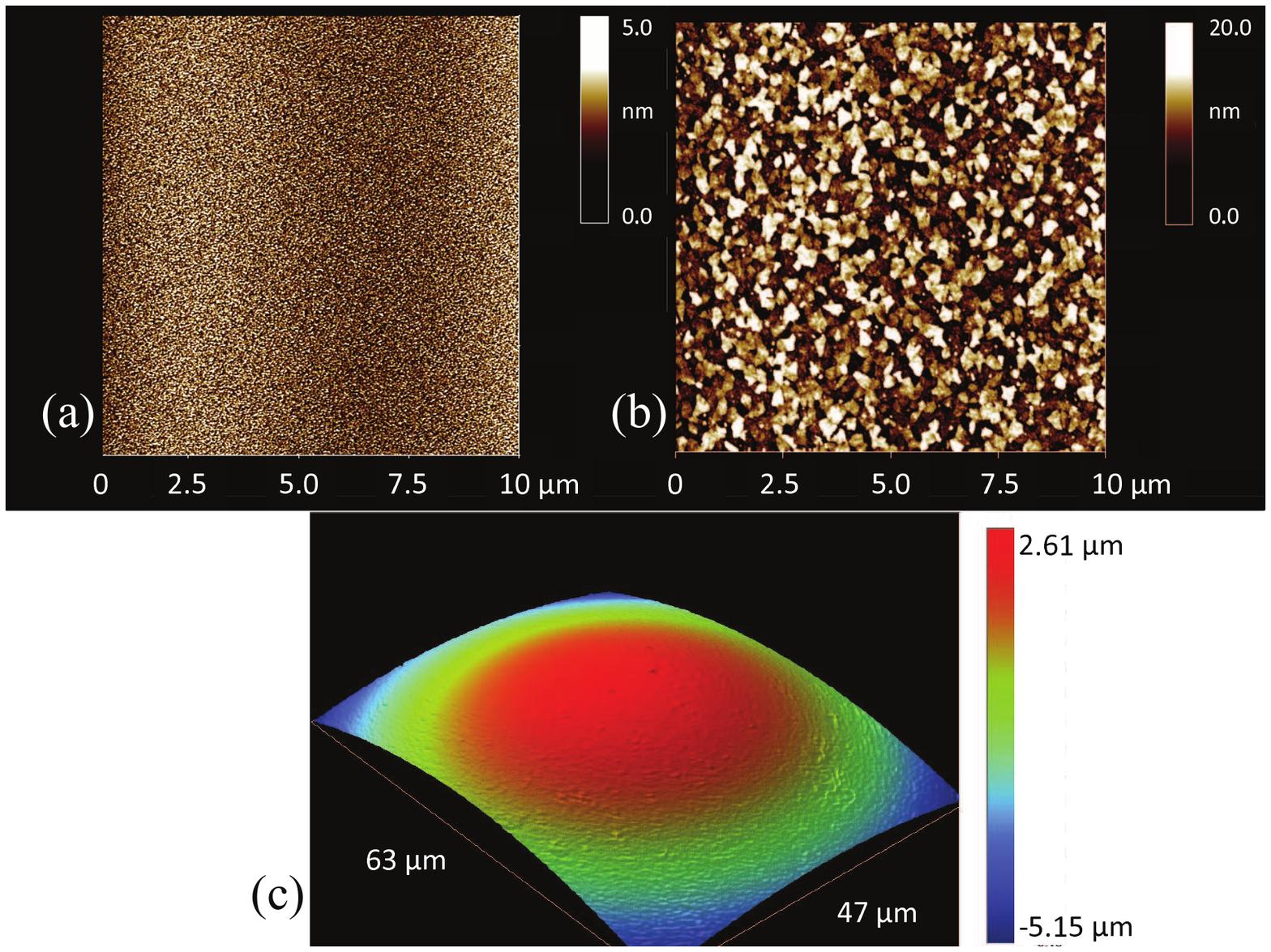}
\caption{(Color online) (a,b) AFM topography scans (10 by 10~$\mu$m) of the surfaces of the Au on sapphire and ITO on float glass samples, respectively. The surface roughness for the gold sample is 0.8~nm RMS, while the ITO plate has a surface roughness of 4~nm RMS. (c) Optical profiler scan of the bottom of the gold coated polystyrene sphere that is attached to the cantilever for our force measurements. The surface roughness of the sphere is 3.8~nm RMS.
}
\label{fig:roughness}
\end{figure}

So far, we have neglected the effects of surface roughness in our analysis. In Fig.~\ref{fig:roughness}, we show topology measurement of our surfaces. Figs.~\ref{fig:roughness}~a and~b are AFM tapping-mode scans (10 by 10~$\mu$m) of the Au on polished sapphire and ITO on float glass samples, respectively. The gold sample has an RMS surface roughness of 0.8~nm, while the ITO coated plate has a roughness of 4~nm RMS. Fig.~\ref{fig:roughness}~c presents a height profile of the surface of the sphere bottom obtained with an optical profiler. Since the cantilever is mounted at a 15 degrees angle with respect to the planar sample surface (this is typical in AFM design), the top of this profile does not correspond to the area of closest approach in a force measurement. However, this height profile does give us the ability to estimate the surface roughness of the sphere, resulting in a value of 3.8~nm RMS.

Since we used the same sphere in both sets of measurements, the surface roughness of the sphere can never cause the observed difference in Casimir force gradients between the Au-Au and Au-ITO cases. Furthermore, we recall that surface roughness tends to enhance the strength of the Casimir interaction~\cite{Maradudin:1980p1905}. It is therefore impossible that the different surface roughnesses of the two planar samples is responsible for the difference reported in Fig.~\ref{fig:ITO}, because the ITO sample is considerably rougher than the gold coated sapphire substrate.

\begin{figure}[!]
\includegraphics[width=13cm]{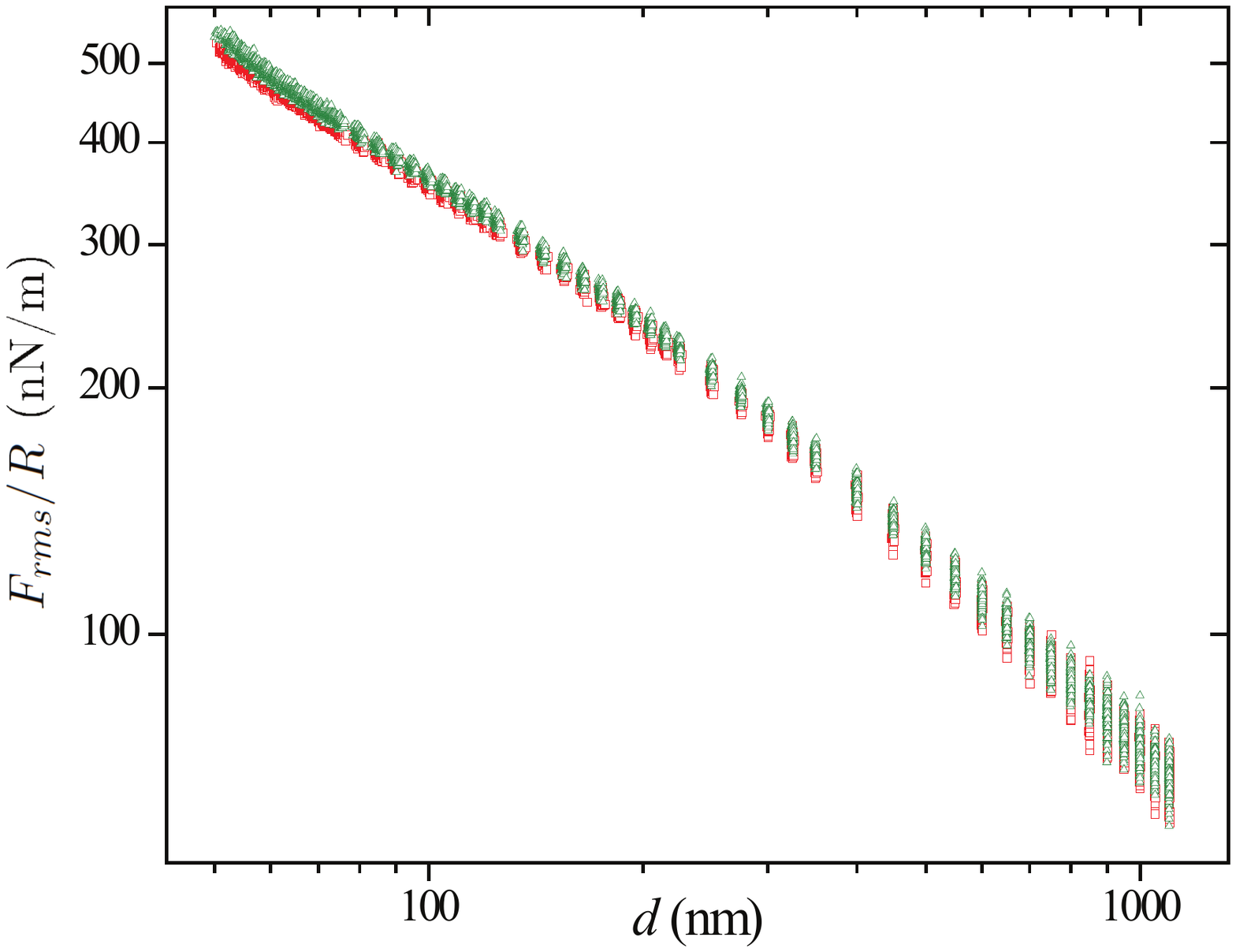}
\caption{(Color online) Hydrodynamic force (RMS) acting on the sphere as a function of separation for a subset of the data (150 out of 580 for both cases), caused by the oscillations of the plate surfaces at 119~Hz. The green triangles represent the force in the case of two gold-coated surfaces, while the red squares
correspond to the Au-ITO interaction.} \label{fig:hydroresults}
\end{figure}

When we discussed the experimental details of our experiment, we mentioned the interesting feature that we can measure both the Casimir force gradient and the hydrodynamic force acting on the sphere with the same lock-in amplifier at $\omega_{2}$. Fig.~\ref{fig:hydroresults} shows the hydrodynamic force for both the Au-Au interaction (green triangles) and for the Au-ITO case (red squares). We have plotted the RMS force resulting from the 2.72~nm RMS oscillation of the plate at $\omega_{2}/2\pi = 119$~Hz. Both curves appear to change exponent at a separation of around 200~nm. This bending is caused by the slip of the air flow across the surfaces, i.~e.~the fluid velocity at the gas-solid interface is nonzero. This phenomenon is treated in~\cite{Hocking:1973p1338} and the expressions derived in there describe our data satisfactorily. Concerning the comparison of the two sets of hydrodynamic data, it is clear that the hydrodynamic forces are very similar in the Au-Au and Au-ITO experiments. Still, there exists a small difference between the two curves of roughly 2\%. This difference cannot be caused by an error in the determination of the initial separation $d_{0}$, because both data sets are parallel on the double logarithmic plot. We suppose that the cause may lie in the different surface roughnesses of the Au and ITO samples that lead to different amounts of fluid slip over the sample surfaces.

It is worthwhile to compare our method for the detection of hydrodynamic forces with recent measurements obtained with AFM's~\cite{Maali:2008p96,Honig:2010p1907}. In~\cite{Maali:2008p96}, the cantilever with sphere is driven at its free resonance and the amplitude and phase of the cantilever motion are used to extract the hydrodynamic force. In~\cite{Honig:2010p1907}, two methods were employed to measure the hydrodynamic interaction between a colloid sphere and a plate: measuring the static deflection of the cantilever during a fast approach of the planar sample and analyzing the thermal noise of the cantilever while slowly approaching the plate towards the sphere. In both papers, however, the separation between the two interacting surfaces was determined by bringing the sphere and plate into contact, a method that is prone to inaccuracies due to surfaces asperities (this is also reported in~\cite{Honig:2010p1907}). Since our method employs both a hydrodynamic force measurement and a precise calibration of the distance at the same time, we have developed a more reliable technique for hydrodynamic force measurements.

\section{Conclusions}
We have presented the experimental details of our Casimir force measurements between gold and ITO surfaces~\cite{deMan:2009p99}. We have shown that the mechanical drift of our setup is less than 0.1~nm per measurement run and that our electrostatic calibration is performed with 0.2\% precision. Force gradient data obtained over approximately 72~hours reveal no drift in the signal at all, confirming the high stability of the setup. Furthermore, we have introduced and demonstrated a new method to determine the spring constant of our cantilever and the deflection sensitivity of the AFM readout. We also presented our measurements of the Casimir and hydrodynamic interactions between the gold and ITO surfaces, and provided
a complete characterization of our samples in terms of their dielectric properties and surface roughness.

\section*{Acknowledgments}
The authors thank R.~J.~Wijngaarden, A.~Baldi, F.~Mul, and J.~H.~Rector for useful discussions. This work was supported by the Netherlands Organisation for Scientific Research (NWO), under the Innovational Research Incentives Scheme VIDI-680-47-209. DI acknowledges financial support from the
European Research Council under the European Community's Seventh Framework Programme
(FP7/2007-2013)/ERC grant agreement number 201739.

\newpage

\end{document}